\input{aipcheck}

\documentclass[
    ,final 
  ] {aipproc}

\layoutstyle{6x9}

\begin{document}

\title{On quark-lepton complementarity}


\author{F. Gonz\'alez Canales}{
  address={Instituto de F\'{\i}sica, UNAM, 04510, M\'exico D.F., MEXICO}  }

\author{ A. Mondrag\'on}{
 address={Instituto de F\'{\i}sica, UNAM, 04510, M\'exico D.F., MEXICO}  }

\begin{abstract}
Recent measurements of the neutrino solar mixing angle and the Cabibbo
angle satisfy the empirical relation $\theta_{sol} + \theta_{C} \simeq
\frac{\pi}{4}$. This relation suggests the existence of a correlation 
between the mixing matrices  of leptons and quarks, the so called  
quark-lepton complementarity.  Here, we examine the possibility that 
this correlation originates in the strong hierarchy in the mass spectra
of quarks  and charged leptons, and  the seesaw mechanism that  
gives  mass to  the Majorana neutrinos.  In a unified treatment of quarks
and leptons in which the mass matrices of all fermions have a similar Fritzsch 
texture, we calculate the mixing matrices $V_{CKM}$ and $U_{MNSP}$ as functions
of quark and lepton masses and only two free parameters, in very good 
agreement with the latest experimental values on masses and mixings. Three 
essential ingredients to explain the quark-lepton complementarity relation are 
identified: the strong hierarchy in the mass spectra of quarks and charged 
leptons, the normal seesaw mechanism and the assumption of maximal CP 
violation in the lepton sector.
\end{abstract}

\maketitle


\section{Introduction}
 In the  last few years, the neutrino oscillations  between different
flavour  states  were  measured   in  a  series  of  experiments  with
atmospheric neutrinos\cite{Super-K}, solar neutrinos\cite{SNO}\cite{SNO05}, 
neutrinos  produced in nuclear  reactors \cite{KamLAND} and accelerators
\cite{acelerador}. As a result, the difference of the squared neutrino masses 
and the mixing  angles   in  the   lepton  mixing  matrix, $U_{MNSP}$, were
determined:
   \begin{equation} 0.34  \leq \sin^{2}{\theta_{23}} \leq  0.68, \qquad
   1.4\times  10^{-3} \textrm{(eV)}^{2}  \leq  \Delta m_{23}^{2}  \leq
   3.0\times 10^{-3} \textrm{(eV)}^{2}, \end{equation}
   \begin{equation}  0.29 \leq \sin^{2}{\theta_{12}}\leq  0.40, \qquad
    7.1  \times 10^{-5}  \textrm{(eV)}^{2}\leq \Delta  m_{12}^{2} \leq
    8.9 \times 10^{-5} \textrm{(eV)}^{2}, \end{equation}
    \begin{equation} \sin^{2}{\theta_{13}}\leq  0.046, \end{equation} 
     at $90 \%$  confidence level \cite{Maltoni}-\cite{Schwetz}. The CHOOZ
     experiment \cite{CHOOZ}    determined   an    upper    bound   for
     the $\theta_{13}$  mixing angle.  It was  soon realized \cite{Smirnov}
     that the solar mixing angle $\theta^{MNSP}_{12}$ and the Cabibbo
     angle  $\theta^{CKM}_{12}$, which is  the corresponding  angle in
     the  quark  sector, satisfiy an interesting  and  intriguing numerical 
relation,
\begin{equation}\label{t12}
 \theta^{MNSP}_{12} + \theta^{CKM}_{12} = 45^{o} + 1^{o} \pm 2.4^{o},
\end{equation} 
with  $\theta_{12}^{MNSP}  =  33.9^{o}  \pm 2.4^{o}$  ($1\sigma$)  and
$\theta_{12}^{CKM}  =  12.8^{o}  \pm 0.15^{o}$.  Equation (\ref{t12})
relates the 1-2 mixing angles in the quark and lepton sectors, it is 
commonly called  the quark-lepton complementarity  relation (QLC) and, 
if not accidental,  it could imply a quark-lepton  symmetry (for a 
recent review see \cite{Minakata}) or a quark-lepton unification 
\cite{Raidal}-\cite{Frampton}.

A second QLC relation, $\theta_{23}^{MNSP}+\theta_{23}^{CKM} \approx
\frac{\pi}{4}$, is also satisfied. However, this is not as interesting as
(\ref{t12}) because $\theta_{23}^{CKM}$ is  only about two degrees, and 
the corresponding QLC relation  would be satisfied, within the errors, 
even  if the angle $\theta_{23}^{CKM}$ had been zero, as long as 
$\theta_{23}^{MNSP}$ is close to the maximal value $\pi/4$. A third 
possible QLC relation  is not realized at  all, or at least not realized 
in the same way, since $\theta_{13}^{CKM}+\theta_{13}^{MNSP}$ is less than 
ten degrees. In this  short note we will focus our attention on understanding 
the nature of the QLC relation shown in equation (\ref{t12}).
\section{Universal Fritzsch texture of quarks and leptons}
The quark and lepton flavour mixing matrices, $U_{MNSP}$ and $V_{CKM}$,
arise from the mismatch between diagonalization of the mass matrices of 
$u$ and $d$ type quarks and the diagonalization of the mass matrices of 
charged leptons and neutrinos,
\begin{equation}\label{M_unitarias}
  U_{MNSP} = U_{l}^{\dagger}U_{\nu}, \quad V_{CKM} = U_{u}^{\dagger}U_{d}.
\end{equation}
Therefore, to get predictions for the flavour mixing angles and CP violating 
phases, we should specify the mass matrices.
 
In this work, we propose a unified treatment of quarks and leptons. Lepton
and quark mass matrices could have the same mass texture from a universal 
flavour symmetry (exact at a certain energy scale). Imposing a flavour 
symmetry has been successful in reducing the number of parameters of the 
Standard Model. In particular, a permutational $S_{3}$ flavour symmetry and its
sequential explicit breaking, allows us to represent the mass matrices as a 
modified Fritzsch texture: 
\begin{equation}\label{T_fritzsch}
  {\bf M^{(F)} = }\left( \begin{array}{ccc} 0 & A & 0 \\ A^{*} & B & C
  \\ 0 & C & D \end{array}\right) \quad i=u,d,l,\nu.
\end{equation}
Some reasons to propose the validity of the modified Fritzsch texture as a 
universal mass texture for all fermions in the theory are the following:
\begin{enumerate}
 \item The idea of $S_{3}$ flavour symmetry and its explicit breaking has 
been realized as a modified Fritzsch texture in the quark sector to 
interpret the strong mass hierarchy of up and down type quarks 
\cite{Fritzsch1}.
\item The quark mixing angles and the CP violating phase appearing in the 
$V_{CKM}$ mixing matrix were computed as explicit, exact functions of the 
four quark mass ratios $(m_{u}/m_{t},m_{c}/m_{t},m_{d}/m_{b},m_{s}/m_{b})$, 
one symmetry breaking parameter $Z^{1/2}=\left(\frac{81}{32}\right)^{1/2}$ 
and one CP violating phase $\phi_{u-d}= 90^{o}$, in very good agreement with 
experiment \cite{MONDRA}.
 \item Since the mass spectrum of the charged leptons exhibits a similar 
  hierarchy to the quark's one, it would be natural to consider the same 
  $S_{3}$ symmetry and its explicit breaking for the charged lepton mass 
matrix.
\item As for the Dirac neutrinos, we have no direct information about the 
absolute values or the relative values of the neutrino masses, but the 
Fritzsch texture can be incorporated in a $SO(10)$ neutrino model 
\cite{W_Wyler}. Therefore it would be sensible to assume that the Dirac 
neutrinos have a mass hierarchy similar to that of the u-quarks and it would 
be natural to take for the Dirac neutrino mass matrix also a modified Fritzsch 
texture.
\item The left handed Majorana neutrinos naturally acquire their mass through 
an effective seesaw mechanism of the form
  \begin{equation}
    M_{\nu_{L}} = M_{\nu_{D}} M_{R}^{-1} M_{\nu_{D}}^{T},
  \end{equation}
where $M_{\nu_{D}}$ and $ M_{R}$ denote the Dirac and right handed Majorana 
neutrino mass matrices. From our conjecture of a universal $S_{3}$ flavour 
symmetry it follows that $M_{R}$ could have the same texture as that of 
$ M_{\nu_{D}}$ and $M_{l}$. Then, it is straightforward to show that 
$M_{\nu_{L}}$ has the same modified Fritzsch texture \cite{Fritzsch}. 
\end{enumerate}

\section{Mixing Matrices as Functions of the Fermion Masses}

When the unitary matrices that diagonalize the mass matrices $M_{i}^{(F)}$ are 
written in polar form, $U_{i} = P_{i} O_{i}$ and $M_{i}^{(F)}=P_{i}^{\dagger}
\bar{M}P_{i}$, the  expressions (\ref{M_unitarias}) for the mixing matrices 
take the form
\begin{equation}\label{M_unitaria2}
 U_{MNSP} = O_{l}^{T}P^{(l -\nu)}O_{\nu}K, \quad V_{CKM} = O_{d}^{T}
 P^{(u-d)}O_{u},
\end{equation}
where $O_{i}$, $i=u, d, \nu, l$, are the orthogonal matrices that diagonalize 
the real symmetric mass matrices $\bar{M}_{i}^{(F)}$ and $P^{(u-d)} =
\textrm{diag}\left[1, e^{i\phi}, e^{i\phi} \right]$, $P^{(l-\nu)} =
\textrm{diag}\left[1, e^{i\Phi}, e^{i\Phi} \right]$, where $\phi = 
\phi_{u}-\phi_{d}$, and $\Phi = \Phi_{l} - \Phi_{\nu}$, are the matrices of 
the Dirac phases and $K$ is the diagonal matrix of the Majorana phases.

We reparametrized the matrices $\bar{M}_{i}$ in terms of their eigenvalues. 
The orthogonal matrices are then expressed in terms of the mass eigenvalues of 
$M_{i}$:
\begin{equation}\label{M_ortogonal}
{\bf O_{i}=}\left(\begin{array}{ccc}
  \left[\frac{\widetilde{m}_{i2}f_{i1}}{D_{i1}}\right]^{\frac{1}{2}}  &
  -\left[\frac{\widetilde{m}_{i1}f_{i2}}{D_{i2}}\right]^{\frac{1}{2}}  &
  \left[\frac{\widetilde{m}_{i1}\widetilde{m}_{i2}f_{i3}}{D_{i3}}\right]
 ^{\frac{1}{2}} \\
  \left[\frac{\widetilde{m}_{i1}(1-\delta_{i})f_{i1}}{D_{i1}}\right]^{\frac{1}
  {2}} & \left[\frac{\widetilde{m}_{i2}(1-\delta_{i})f_{i2}}{D_{i2}}\right]
  ^{\frac{1}{2}}
  &  \left[\frac{(1-\delta_{i})f_{i3}}{D_{i3}}\right]^{\frac{1}{2}} \\
  -\left[\frac{\widetilde{m}_{i1}f_{i2}f_{i3}}{D_{i1}}\right]^{\frac{1}{2}}
  &
  -\left[\frac{\widetilde{m}_{i2}f_{i1}f_{i3}}{D_{i2}}\right]^{\frac{1}{2}}
  &  \left[\frac{f_{i1}f_{i2}}{D_{i3}}\right]^{\frac{1}{2}}
\end{array}\right) ,
\end{equation}
\begin{equation}\label{fs}
  f_{i1}=1-\widetilde{m}_{i1}-\delta_{i}, \quad  f_{i2} =1+ \widetilde{m}_{i2}
  -\delta_{i}, \quad f_{i3}=\delta_{i},
\end{equation}
\begin{equation}
 D_{i1} = (1 - \delta_{i})(\widetilde{m}_{i1} + \widetilde{m}_{i2} ) ( 1- 
  \widetilde{m}_{i1}),  
\end{equation}
\begin{equation}
  D_{i2} = (1 - \delta_{i})(\widetilde{m}_{i1} + \widetilde{m}_{i2} ) ( 1+
  \widetilde{m}_{i2}),
\end{equation}
\begin{equation}\label{D3}
 D_{i3} = (1 - \delta_{i})(1-\widetilde{m}_{i1}) ( 1+\widetilde{m}_{i2}),
\end{equation}
the small parameters $\delta_{i}$ are also functions of the mass ratios and the
symmetry breaking parameter $Z^{1/2} = (81/32)^{1/2}$. 

Substitution of the expressions (\ref{M_ortogonal}) and (\ref{fs})-(\ref{D3}) 
in (\ref{M_unitaria2}) allows us to express the mixing matrices $U_{MNSP}$ and 
$V_{CKM}$ as explicit functions of the quark and lepton masses.
\section{Quark-Lepton Complementarity}

The resulting theoretical expression for the Cabibbo angle written to first 
order in $m_{u}/m_{c}$ and $m_{d}/m_{s}$, is
\begin{equation}
 \sin^{2}{\theta_{C}^{\textrm{th}}} = \mid V_{us}^{\textrm{th}}
 \mid^{2} \approx  \frac{ \frac{\widetilde{m}_{d}}{\widetilde{m}_{s}}
 + \frac{\widetilde{m}_{u}}{\widetilde{m}_{c}} - 2
 \sqrt{\frac{\widetilde{m}_{u}}{\widetilde{m}_{c}}
 \frac{\widetilde{m}_{d}}{\widetilde{m}_{s}}} \cos{\phi} }{  \left( 1
 +  \frac{\widetilde{m}_{u}}{\widetilde{m}_{c}}\right)\left( 1 +
 \frac{\widetilde{m}_{d}} {\widetilde{m}_{s}}  \right)}.
\end{equation}
Taking for the quark masses the values $m_{u}=2.75$MeV, $m_{c}=1310$MeV, 
$m_{d}=6.0$MeV, $m_{s}=120$MeV and maximal CP violation, $\phi=90^{o}$
\cite{MONDRA}, we reproduce the numerical value of the Cabibbo angle 
\begin{equation}
\sin{\theta_{c}^{\textrm{th}}} = 0.225 \quad \textrm{or} \quad 
\theta_{c} = 12.8^{o},
\end{equation}
in very good agreement with the latest analysis of the experimental data
\cite{mateu}.

The theoretical expression for the solar mixing angle is derived in a similar 
way. From $\mid (U_{MNSP})_{12} \mid^{2} / \mid (U_{MNSP})_{11} \mid^{2}= 
\tan^{2}{\theta_{12}^{\textrm{th}}}$, we obtain
\begin{equation}
\tan^{2}{\theta_{12}^{\textrm{th}}} =
  \frac{\frac{\widetilde{m}_{\nu_{1}}}{\widetilde{m}_{\nu_{2}}} +
  \frac{\widetilde{m}_{e}}{\widetilde{m}_{\mu}} -
  2\sqrt{\frac{\widetilde{m}_{\nu_{1}}}{\widetilde{m}_{\nu_{2}}}
  \frac{\widetilde{m}_{e}}{\widetilde{m}_{\mu}}} \cos{\Phi}}{ 1 +
  \frac{\widetilde{m}_{\nu_{1}}} {\widetilde{m}_{\nu_{2}}}
  \frac{\widetilde{m}_{e}}{\widetilde{m}_{\mu}} +
  2\sqrt{\frac{\widetilde{m}_{\nu_{1}}}{\widetilde{m}_{\nu_{2}}}
  \frac{\widetilde{m}_{e}}{\widetilde{m}_{\mu}}} \cos{\Phi} }.
\end{equation}
In the absence of experimental information, we assumed that CP violation is 
also maximal in the lepton sector $i.e.$ $\Phi=90^{o}$. Taking for the masses 
of the left handed Majorana neutrinos a normal hierarchy with the numerical 
values $m_{\nu_{1}} = 4.4 \times 10^{-3}$eV and $m_{\nu_{2}} = 1.0 \times 
10^{-2}$eV, and for the charged lepton masses the values $m_{e} = 0.5109$MeV, 
$m_{\mu}= 105.685$MeV and $m_{\tau}=1776.99$GeV, we obtain the following 
numerical value for the solar mixing angle
\begin{equation}
\tan^{2}{\theta_{12}^{\textrm{th}}} = 0.45 \quad \textrm{or} \quad
\theta_{12}^{\textrm{th}} = 33.9^{o}.
\end{equation}
  
We may now address the question of the meaning of the quark-lepton 
complementarity relation as expressed in eq(\ref{t12}). The previous 
theoretical analysed allows us to calculate,
\begin{equation}
 \tan{\left( \theta_{c}^{\textrm{th}} + \theta_{12}^{\textrm{th}}\right)} = 1 +
\Delta^{\textrm{th}},
\end{equation}
{\footnotesize
\begin{equation}\label{DELTA}
 \Delta^{\textrm{th}} = \frac{\left( \frac{\widetilde{m}_{d}}{\widetilde{m}_{s}
}+\frac{\widetilde{m}_{u}}{\widetilde{m}_{c}}\right)^{\frac{1}{2}}\left[ \left(
\frac{\widetilde{m}_{\nu_1}}{\widetilde{m}_{\nu_2}} + \frac{\widetilde{m}_{e}}{
\widetilde{m}_{\mu}}\right)^{\frac{1}{2}} + \left( 1 + \frac{\widetilde{m}_{
\nu_1}}{\widetilde{m}_{\nu_2}}\frac{\widetilde{m}_{e}}{\widetilde{m}_{\mu}}
\right)^{\frac{1}{2}}\right] + \left( 1 + \frac{\widetilde{m}_{d}}{\widetilde{m
}_{s}}\frac{\widetilde{m}_{u}}{\widetilde{m}_{c}} \right)^{\frac{1}{2}} \left[ 
\left(\frac{\widetilde{m}_{\nu_1}}{\widetilde{m}_{\nu_2}} + \frac{\widetilde{m
}_{e}}{\widetilde{m}_{\mu}}\right)^{\frac{1}{2}} - \left( 1 + \frac{\widetilde{
m}_{\nu_1}}{\widetilde{m}_{\nu_2}}\frac{\widetilde{m}_{e}}{\widetilde{m}_{\mu}}
\right)^{\frac{1}{2}} \right]}{\left( 1 + \frac{\widetilde{m}_{\nu_1}}{
\widetilde{m}_{\nu_2}}\frac{\widetilde{m}_{e}}{\widetilde{m}_{\mu}}\right)^{
\frac{1}{2}}\left( 1 + \frac{\widetilde{m}_{d}}{\widetilde{m}_{s}}\frac{
\widetilde{m}_{u}}{\widetilde{m}_{c}} \right)^{\frac{1}{2}} - \left( \frac{
\widetilde{m}_{d}}{\widetilde{m}_{s}} + \frac{\widetilde{m}_{u}}{\widetilde{m}
_{c}}\right)^{\frac{1}{2}}\left(\frac{\widetilde{m}_{\nu_1}}{\widetilde{m}
_{\nu_2}} + \frac{\widetilde{m}_{e}}{\widetilde{m}_{\mu}}\right)^{\frac{1}{2}}}
.
\end{equation}}
After substitution of the numerical value of the mass ratios of quarks and 
leptons in (\ref{DELTA}), we obtain,
\begin{equation}
 \Delta^{\textrm{th}} = 0.061 \quad \textrm{and} \quad \theta_{c}^{\textrm{th}}
 + \theta_{12}^{\textrm{th}} = 45^{o} + 1.7^{o}. 
\end{equation}
in very good agreement with the experimental value.
\section{CONCLUSIONS}
In this short communication, we outlined a unified treatment of masses and 
mixing of quarks and leptons in which the left handed Majorana neutrinos 
acquire their masses via the seesaw mechanism, and the mass matrices of all 
fermions have a similar Fritzsch texture and a normal hierarchy. In this 
scheme, we derived exact, explicit expressions for the Cabibbo and solar 
mixing angles as functions of the quark and lepton masses. The quark-lepton 
complementarity  relation takes the form,
\begin{equation}
 \theta_{12}^{\textrm{CKM}} + \theta_{12}^{\textrm{MNSP}} = 45^{o} + 
\delta_{12}.
\end{equation}
The correction term, $\delta_{12}$, is an explicit function of the ratios of 
quark and lepton masses, given in eq.(\ref{DELTA}), which reproduces the 
experimentally determined value, $\delta_{12} \approx 1.7^{o}$, when the 
numerical values of the quark and lepton masses are substituted in 
(\ref{DELTA}) and maximal violation of CP in the lepton sector is assumed.

Three essential ingredients are needed to explain the correlations implicit in 
the small numerical value of $\delta_{12}$:
\begin{enumerate}
\item The strong hierarchy in the mass spectra of the quarks and charged 
leptons, realized in our scheme through the explicit breaking of the $S_{3}$ 
flavour symmetry in the Fritzsch mass texture, explains the resulting small or
 very small quark mixing angles, the very small charged lepton mass ratios 
explain the very small $\theta_{13}^{MNSP}$ which, in our scheme, is 
independent of the neutrino masses.
\item The normal seesaw mechanism that gives very small masses to the left 
handed Majorana neutrinos with relatively large values of the neutrino mass 
ratio $m_{\nu_1}/m_{\nu_{2}}$ and allows for large 
$\theta_{12}^{\textrm{MNSP}}$ and $\theta_{23}^{\textrm{MNSP}}$ mixing angles.
\item The assumption of maximal CP violation in the lepton sector.
\end{enumerate}

A more complete and detailed version of this work will be presented in a 
forthcoming publication \cite{FyA}
\section{Acknowledgements}
We thank Dr. M. Mondrag\'on for many inspiring discussions on this exciting 
problem and for a critical reading of the manuscript.

This work was partially supported by CONACyT Mexico under Contract No. 42026F, 
and DGAPA-UNAM Contract No. PAPIIT IN116202.

\end{document}